\documentclass[aps,a4paper,floatfix,nofootinbib]{revtex4}

\usepackage{graphicx,color}
\usepackage{amsmath,amssymb,amsfonts,amsthm,amscd,bm}
\usepackage{booktabs}
\usepackage{cancel}

\def\tilde{\widetilde}
\def\bar{\overline}
\def\hat{\widehat}
\def\*{\star}
\def\[{\left[}
\def\]{\right]}
\def\({\left(}      

\def\){\right)}

\def\frac#1#2{\dfrac{#1}{#2}}
\def\inv#1{\dfrac{1}{#1}}
\def\half{\tfrac{1}{2}}
\def\d{\partial}

\def\2pi{\hbox{$2\pi i$}}

\def\dsl{\raise.15ex\hbox{/}\kern-.57em\partial}
\def\Dsl{\,\raise.15ex\hbox{/}\mkern-.13.5mu D}
\def\CD{{\cal D}}   \def\CE{{\cal E}}

      \def\CO{{\cal O}}
      
\def\CS{{\cal S}}

\def\2pi{\hbox{$2\pi i$}}

\def\dsl{\raise.15ex\hbox{/}\kern-.57em\partial}
\def\Dsl{\,\raise.15ex\hbox{/}\mkern-.13.5mu D}
\font\numbers=cmss12
\font\upright=cmu10 scaled\magstep1
\def\stroke{\vrule height8pt width0.4pt depth-0.1pt}
\def\topfleck{\vrule height8pt width0.5pt depth-5.9pt}
\def\botfleck{\vrule height2pt width0.5pt depth0.1pt}
\def\Zmath{\vcenter{\hbox{\numbers\rlap{\rlap{Z}\kern
    0.8pt\topfleck}\kern 2.2pt
    \rlap Z\kern 6pt\botfleck\kern 1pt}}}
\def\Qmath{
    \vcenter{\hbox{\upright\rlap{\rlap{Q}\kern3.8pt\stroke}\phantom{Q}}}}
\def\Nmath{\vcenter{\hbox{\upright\rlap{I}\kern 1.7pt N}}}
\def\Cmath{\vcenter{\hbox{\upright\rlap{\rlap{C}\kern
                   3.8pt\stroke}\phantom{C}}}}
\def\Rmath{\vcenter{\hbox{\upright\rlap{I}\kern 1.7pt R}}}
\def\Z{\ifmmode\Zmath\else$\Zmath$\fi}
\def\Q{\ifmmode\Qmath\else$\Qmath$\fi}
\def\N{\ifmmode\Nmath\else$\Nmath$\fi}
\def\C{\ifmmode\Cmath\else$\Cmath$\fi}
\def\R{\ifmmode\Rmath\else$\Rmath$\fi}
\def\barray{\begin{eqnarray}}
\def\earray{\end{eqnarray}}
\def\beq{\begin{equation}}
\def\eeq{\end{equation}}

\def\AA{\leavevmode\setbox0=\hbox{h}
\dimen0=\ht0 \advance\dimen0 by-1ex\rlap{\raise.67\dimen0\hbox{\char'27}}A}
\def\chitilde{\hat{\chi}}

\def\Arg{{\rm Arg}\,}

\makeatletter
\def\iddots{\mathinner{\mkern1mu\raise\p@
\vbox{\kern7\p@\hbox{.}}\mkern2mu
\raise4\p@\hbox{.}\mkern2mu\raise7\p@\hbox{.}\mkern1mu}}
\makeatother

\theoremstyle{plain}

\theoremstyle{remark}

\def\arg{{\rm arg}}
\def\Arg{{\rm Arg}}
\def\half{\tfrac{1}{2}}

\def\rhozero{\rho_\bullet}

\def\tzero{t_\bullet}
\def\thetaRS{\vartheta}
\def\ttilde{{\tilde{t}}}

\def\ttilde{\tilde{t}}

\def\charphi{\varphi} 

\def\character{X}

\def\r{r} 

\def\q{q} 
\def\xi{\chi}
\def\Soft{\CS}
\renewcommand{\thefootnote}{\roman{footnote}} 

\begin{document}

\title{
Riemann zeros as quantized energies of  scattering with impurities }
\author{Andr\'e  LeClair\footnote{Andre.LeClair@cornell.edu \\ mussardo@sissa.it}}
\affiliation{Cornell University, Physics Department, Ithaca, NY 14850, USA} 
\author{Giuseppe Mussardo}
\affiliation{SISSA and INFN, Sezione di Trieste, via Bonomea 265, I-34136, 
Trieste, Italy}
\renewcommand{\thefootnote}{\alph{footnote}}

\begin{abstract}

We construct an integrable physical model of a single particle scattering with impurities spread on a circle. The $S$-matrices of the scattering with the impurities are such that the quantized energies of this system, coming from the Bethe Ansatz equations, correspond to the imaginary parts of the non-trivial zeros of the the Riemann $\zeta(s) $ function along the axis $\Re ( s )= \half$ 
of the complex $s$-plane. A simple and natural generalization of the original scattering problem leads instead to Bethe Ansatz equations whose solutions are the non-trivial zeros of the Dirichlet $L$-functions again along the axis $\Re (s)  = \half$.
The conjecture that all the non-trivial zeros of these functions are aligned along this axis of the complex $s$-plane is known as the Generalised Riemann Hypothesis (GRH). In the language of the scattering problem analysed in this paper the validity of the GRH is equivalent to the completeness of the Bethe Ansatz equations. Moreover the idea that the validity of the GRH requires both the duality equation (i.e. the mapping $s \rightarrow 1 - s$) and the Euler product representation of the Dirichlet $L$-functions finds additional and novel support from the physical scattering model analysed in this paper. 
This is further illustrated by an explicit counterexample provided by the solutions of the Bethe Ansatz equations which employ the Davenport-Heilbronn function $\CD (s) $, i.e. a function whose completion  satisfies the duality equation $\chi(s) = \chi(1-s)$ but that does not have an Euler product representation. In this case, even though there are infinitely many solutions of the Bethe Ansatz equations along the axis $\Re (s)  = \half$, 
there are also infinitely many pairs of solutions away from this axis and symmetrically placed with respect to it.

\end{abstract}

\maketitle

\section{Introduction}

If the Riemann Hypothesis has fascinated mathematicians for decades (see \cite{Edwards,Titchmarsh,Borwein,Conrey,Apostol}), it has also fascinated theoretical physicists for quite a long time (for a review up to 2011, see for instance \cite{Schumayer}). The idea that a remarkable mathematical property such as the alignment of the infinite number of zeros of the Riemann zeta function $\zeta(s)$ along the axis $\Re(s) = \half$ 
may be understood from the simple and elegant requirements of a physical system is too appealing to pass up. ``Understanding" is obviously different from ``proving” but it may nevertheless be the first promising step toward a more rigorous approach. It is precisely with such a  ``theoretical physicist attitude" that we approach the famous problem of the alignment  of all zeros of the Riemann zeta function $\zeta(s)$ along the axis 
$\Re (s) = \half$.

One of the most prominent physical proposals of the past is probably the Hilbert-P\'olya idea which turns the problem of establishing the validity of the Riemann Hypothesis into the existence of a single particle quantum Hamiltonian whose eigenvalues are equal to the ordinates of zeros on the critical line. This has been pursued in several relevant papers, such as  \cite{BerryK,BerryK2,Sierra,Sredincki,Bender}. In that approach, one searches for a quantum hamiltonian where the {\it bound state} energies are the non-trivial Riemann zeros on the critical line. Such a hypothetical hamiltonian remains unknown, and although it is looking more and more unlikely that there exists a closed and simple formula for it, it still remains a possibility. It is worth mentioning that the zeta function has also been employed in describing scattering amplitudes in quantum mechanics \cite{Gutzwiller} and quantum field theory \cite{Remmen}.

 A different line of attack, based on statistical physics and random walks, has recently been pursued \cite{GRHstoch,LMDirichlet,MLRW,FL2}. The aleatory nature of the problem arises from the pseudo-randomness of  Dirichlet characters over the prime numbers, or the M\"{o}bius coefficients evaluated on primes and, nicely enough, this property can be checked with astonishing accuracy \cite{MLRW}. Up to logarithmic  corrections, in this statistical physics approach the validity of the Riemann Hypothesis can be easily understood in physical terms from the universality of the critical exponent $1/2$ of the random walk.

In this article we define a new approach: formulated as a quantum mechanical scattering problem, it has undoubtedly a theoretical physics origin and therefore can be of interest both for theoretical physicists and mathematicians. Our proposal radically differs from those previously mentioned \cite{Gutzwiller,Remmen} for, in the model we construct, there is a quantization condition for the energies of the system which comes from a Bethe Ansatz equation, i.e. the solution of the model employs the same successful formalism used in the past to study a wide spectrum of phenomena, from quantum spin chains (see for instance \cite{Sutherland,Gaudin} and references therein) to AdS/CFT correspondence (see for instance, \cite{AdSCFT} and references therein). The Bethe Ansatz permits to determine self-consistently the energy levels of quantum systems which possess infinitely many conserved quantities, i.e. are integrable. In our case, the solutions of the Bethe Ansatz equation are {\it exactly} the Riemann zeros.  It is interesting to underscore  that, in our formulation, the validity of the Riemann Hypothesis becomes equivalent to the physical condition of {\em completeness} of the Bethe Ansatz solutions, i.e. the condition that guarantees the existence and the uniqueness  of a real solution for any value of the quantization integer $n$ entering the Bethe Ansatz equations, and ensures that the Bethe Ansatz wave functions form a basis for square integrable functions on a circle.

The considerations presented above for the Riemann $\zeta$-function turn out to be also true for the more general class of the  L-functions
based on Dirichlet characters, for which the Generalised Riemann Hypothesis is assumed to hold: indeed, as the Riemann $\zeta$-function, these functions also enjoy a duality condition and admit an infinite product representation\footnote{For the definition of the Dirichlet $L$-functions, see eq.\,(\ref{EPFchi}) below, while for their main properties see for instance ref. \cite{Apostol}. Here it is sufficient to remind that, given a positive integer $q$, these functions are Dirichlet series whose coefficients are the characters of the abelian group $({\mathbb Z}/q\, {\mathbb Z})$ of prime residue classes modulo q. The elements of this group are the integers $a$ such that $a \,{\rm mod}\, q : {\rm gcd}(a, q) = 1$,  The Riemann $\zeta$-function corresponds to the Dirichlet $L$-function with $q=1$. For a given $q$, the principal character $\character(n)$ has the following values $\character(n) = \left\{\begin{array}{ccc} 1 &,& {\rm if} \, {\rm gcd}(n, q) = 1 \\ 0 &,& {\rm otherwise} \end{array} \right.$.}
.
As a matter of fact, the Riemann $\zeta$-function is just a particular case of the Dirichlet $L$-function and, for this reason, unless explicitly stated, when we refer in the following to the Dirichlet $L$-functions we are implicitly referring to the Riemann $\zeta$-function too.

Let us also explicitly stress that there is an additional and important motivation for involving in our discussion the more general case of the Dirichlet $L$-functions. Indeed, as previously shown in \cite{GRHstoch,LMDirichlet,MLRW}, dealing with Dirichlet L-functions of non-principal characters often proves to be technically easier than dealing with the Riemann $\zeta$ function (as well as with any other Dirichlet L-functions of principal characters): the reason is that the Dirichlet L-functions of non-principal characters are {\em entire} functions in the complex plane  due to the absence of the pole at $s=1$. This fact guarantees better convergence properties of many of their related quantities, as we are also going to see in the Bethe Ansatz approach pursued in this paper. For the Bethe Ansatz equations relative to all Dirichlet L-functions, the solutions are {\it exactly} the non-trivial zeros of these functions aligned along the axis $\Re (s) = \half$.  This result relies in an essential manner on both the functional equation {\it and}  the Euler infinite product representation of these functions. In a nutshell, the reason why {\em both} these two conditions are important is the following: duality, i.e. the mapping $s \rightarrow 1 - s$, selects the line $\Re (s) = \half$ as the privileged axis of the Dirichlet functions while their infinite products and the possibility to truncate them to any desired order (with nice properties of convergence of the corresponding truncated expressions) guarantees instead the continuity of the Bethe Ansatz equations, i.e. the possibility to always find a complete set of real solutions.

In order to obtain the right perspective on the  important remark  behind the validity of the GRH of the last paragraph, in the following we discuss an explicit counter-example, namely the case of Bethe Ansatz equations based on scattering theory which employs the Davenport-Heilbronn $L$-function. This function satisfies a duality equation but it does not have an infinite Euler product representation. In this case we will explicitly show that the Bethe Ansatz solutions are {\em not} complete, namely the Bethe Ansatz equations may be discontinuous and do {\em not} always admit real solution for all values of the quantization integer $n$. This counter-example enlightens the important role played by the infinite product representation of the Riemann and Dirichlet L-function for establishing the validity of the Generalised Riemann Hypothesis.

The paper is organised as follows: in Section II we are concerned with the general setting of elastic scattering with $N$ impurities spread on a circle that leads to the Bethe Ansatz equations. In Section III we employ expressions of the $S$-matrix of the impurities which, asymptotically in the energy, give rise to Bethe Ansatz equations for the non-trivial zeros of the Riemann $\zeta$-function. In this section we also discuss the delicate issue of how to take the thermodynamic  limit $N \rightarrow \infty$ of the Bethe ansatz equations. As we will see, one of the options is to turn the attention to Dirichlet $L$-functions with non-principal characters since those functions are expected to have nicer convergent expressions for any of their relevant quantities. In Section IV we refine the discussion of the previous section, showing how to arrive to Bethe Ansatz equations whose solutions are expected to be {\em exactly} the zeros of the Riemann or of the Dirichlet function along the critical line. In Section V we recall  the main mathematical properties of the phase-shift coming directly from the Riemann $\zeta$ function and their statistical properties dictated by a gaussian distribution. In Section VI we present the useful definition of ``regular alternating function" and discuss some other extra mathematical properties of the completed  Riemann $\zeta$ function along the critical line, namely its remarkable Fourier transform and its zero mean. 
In Section VII we present the counter-example of the Davenport-Heilbronn function (i.e. a function invariant under duality but without an infinite product representation) 
which leads to Bethe Ansatz equations with complex solutions, i.e. with zeros away from the critical axis. Finally our conclusions can be found in Section VIII.

\def\p{p}  
\def\pprime{\mathfrak{p}}
\def\r{r}

\section{Bethe ansatz equation for impurities on a circle:  The most  general case}

\begin{figure}[t]
\centering\includegraphics[width=.7\textwidth]{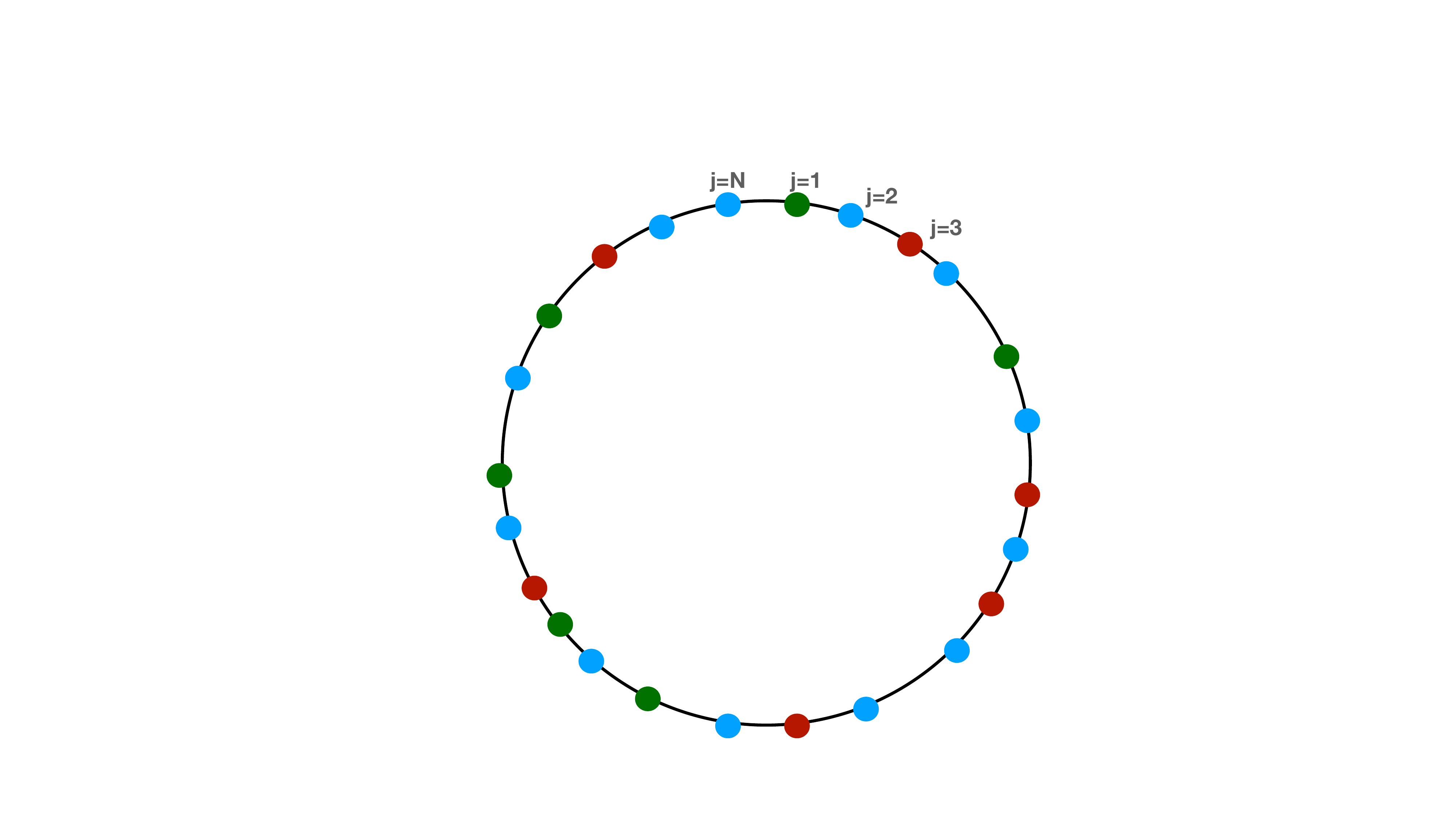}
\caption{Impurities  on a circle labeled $j=1, 2,..., N$.   Different colors denote different scattering phases.   There are not enough colors to represent the scattering problem considered here.  }
 \label{DefectFigure}
\end{figure}

Consider a single particle of momentum $\p$ moving on a circle of circumference $R$ without any internal degree of freedom. Such a particle has a dispersion relation $E(p)$ where $E$ is the energy of the particle,  typically relativistic or non-relativistic. For  generality we leave this dispersion relation  unspecified for this section. We suppose there are $N$ stationary impurities spread out on the circle,  with no particular location,  except that they are separated,  and label them $j = 1, 2, \ldots N$,  as illustrated in Figure \ref{DefectFigure}. When the particle  scatters through a single impurity,  there is generally both a transmission and reflection amplitude.    We assume there is no reflection,   namely the scattering is purely transmitting. There are many known examples of  purely transmitting relativistic theories \cite{DMS,KonikLeClair,Corrigan},  in fact infinitely many that are integrable, and there are also non-relativistic examples of reflectionless potentials \cite{NonRelativistic0,NonRelativistic}.  
To each impurity labeled $j$ we associate a transmission S-matrix $S_j (\p)$,  which by unitarity, is a phase
\beq
\label{Si}
S_j (\p ) = e^{i \phi_j (\p)} \,\,\,.
\eeq
Due to the purely transmitting property,  the  scattering  matrix for 2 impurities $j,j'$ is simply  $S_j (\p) S_{j'}(\p) $,  and so on.

As the particle moves around the circle, it scatters through each impurity and, coming back to its original position, the matching requirement for its wavefunction leads to the quantization condition of its momentum $p$ expressed by
\beq
\label{Bethe1} 
e^{i \p R} \, \prod_{j=1}^N  S_j (\p) = \pm 1 \,\,\,,
\eeq
where $+1, -1$ corresponds to bosons, fermions respectively (see, for instance \cite{MussardoBook} and references therein).   
If we take the particles to be fermions, we end up with the trascendental Bethe-ansatz equation \cite{yangyang,ZamTBA}
\beq
\label{Bethe2}
\p_n  R +  \sum_{j=1}^N   \phi_j (\p_n  ) = 2 \pi  \, (n - \half) \,,
\eeq
for some integer $n$.   Then the quantized energies of the system are $E_n = E(p_n)$, where $p_n$ is the solution of the Bethe-ansatz equation relative to the integer $n$. 

There are several physical applications of this general formula: let us mention, for instance, that if the scattering phases $\phi_j$ are random, this is essentially a problem of
electrons in a random potential in $1$ dimension and related to Anderson localization.

\bigskip

\section{Scattering problem that asymptotically yields the Riemann and Dirichlet zeros on the critical line}

In this section we construct a physical scattering problem where the  quantized $E_n$  of the last section are asymptotically the Riemann  zeros on the critical line.
By ``asymptotically" we mean  large $E_n$  but, as we will explain, very low $n$ (namely $n>  2$) is already asymptotic enough,   however we will still refer to this as ``asymptotic.     Later we generalize the considerations to the Dirichlet $L$-functions.   

\subsection{Dispersion relation}

\begin{figure}[t]
\centering\includegraphics[width=.5\textwidth]{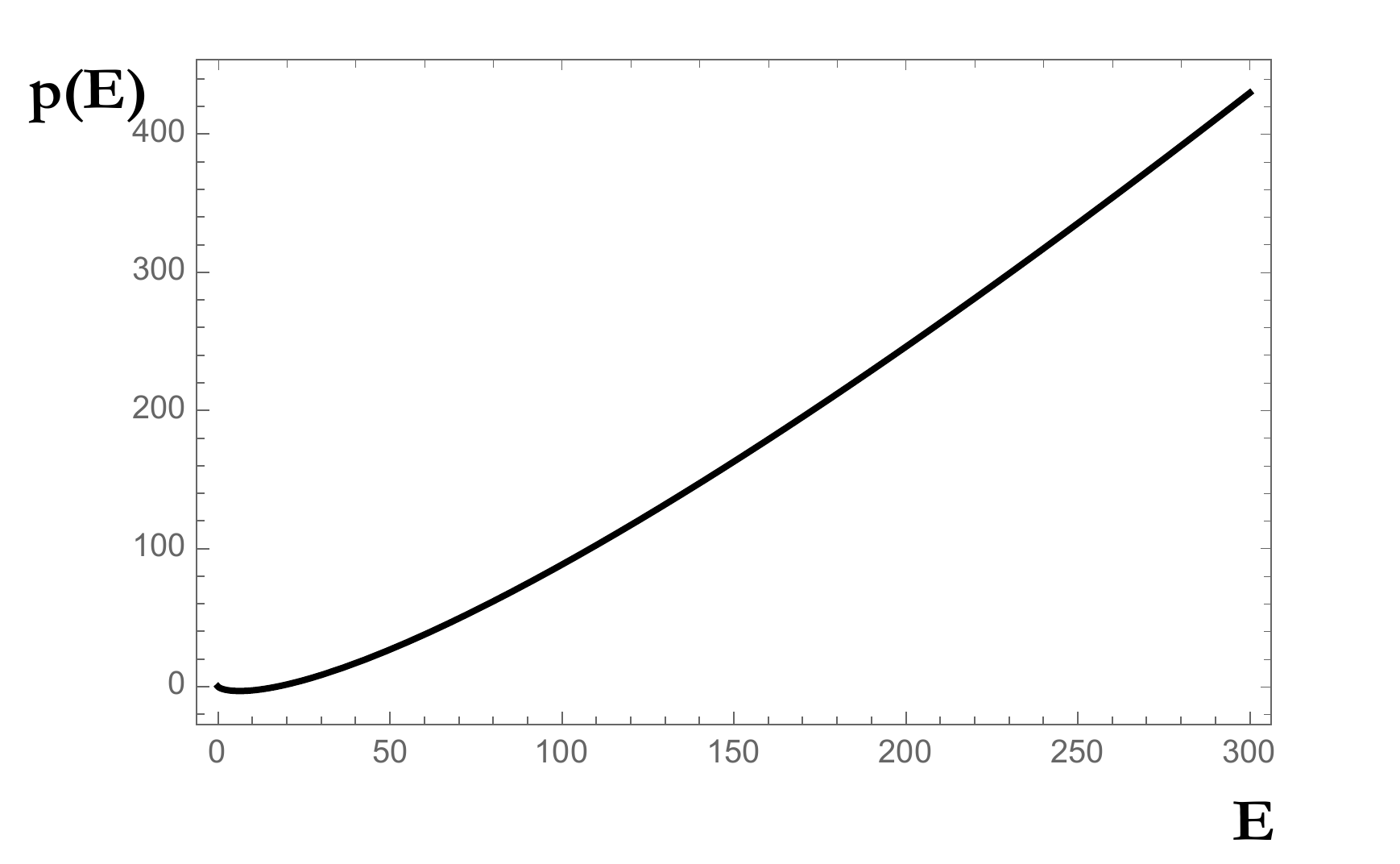}
\caption{Dispersion relation $p(E)$ for our scattering problem. For $E > 2 \pi e$, the only values we are interested in, the function is monotonic and invertible.  }
 \label{dispersion}
\end{figure}

Let's first specify the dispersion relation $E(\p)$ for the free particle between impurities for which we take 
\beq
\label{Dispersion}
\p (E) = E \log (E/\hbar \omega) / v
\eeq
where $v$ is a speed, such as the speed of light, and $\omega$ is a fixed  frequency with units of inverse time. Without the $\log E$ factor, this is a relativistic dispersion relation for a massless particle where $v$ is the speed of light.  We henceforth  set $\hbar =1$. Without loss of generality we can redefine $v, \omega$ such that $\p$ and $E$ are dimensionless,  and 
\beq
\label{Dispersion2}
\p (E)  = E \, \log \( \frac{E}{2 \pi  e } \) 
\eeq
where $e$ is the Euler number, i.e. $1 = \log e$. Note that $\p$ is a monotonic function of $E$ for $E \geq 2 \pi$ and positive\footnote{In the following 
we are interested in values of $E \geq 2 \pi e$ for matching the request of a physical dispersion relation for the momentum $p$ as a function of the energy $E$.
However, as we see later, the value $n=1$ in eq.\,(\ref{FLasymptotic}) makes the RHS of this equation negative. Although the corresponding solution $E \simeq 14.1347...$ 
is within the interval where $p(E)$ is negative, it is however remarkably close to the first non-trivial zero of the Riemann $\zeta$-function along the axis $\Re ( s )= \half$.} for 
$E > 2 \pi e$  (see figure \ref{dispersion}).  Hence, in the infinite interval for $E > 2 \pi e$ the above dispersion relation can be inverted as 
\beq
\label{Eofp}
E(\p)  = \frac{p}{ W(\p/2 \pi  e )}
\eeq
where $W$ is the principal branch of the Lambert $W$ function. 
For large $x$, 
$W(x) = \log x - \log \log x + ....$.
Thus for large $p$, 
\beq
\label{Elargep}
E(\p) \approx  \frac{\p}{\log \p} .
\eeq
Henceforth when we say ``monotonic" it is implicit that this is for $E$ above the low value $ 2 \pi e$ where the function is also positive.

\def\r{\pprime} 
\def\p{\pprime}

\subsection{Scattering phases} 
Let us now specify the scattering phases.    We suppose that the transmission S-matrices are more easily expressed in terms of the energy $E$ rather than  the momentum $p$.   
To each impurity $j$ we associate a positive real number $\q_j >1$ and a constant  real phase angle $\charphi_j$:
\beq
\label{Sjs}
S_j (E) = \frac{\q_j^{\sigma}    - e^{i (E \log \q_j - \charphi_j )}  }
{ \q_j^{\sigma}   - e^{- i (E \log \q_j  - \charphi_j )}   } ,
\eeq
where $\sigma$ is a free parameter which is a  positive real number.     
 This implies that the scattering phases are 
\beq
\label{phis}
\phi_j (E) =  - 2 \, \Im \log \(  1 -  \frac{e^{- i (E \log \q_j  - \charphi_j )} } { \q_j^{\sigma} }  \)\, .
\eeq
Without loss of generality, we set $R=1$, since  $R$  with physical dimension of length can be absorbed into various constants above, such as $v, \omega$.   The impurities do not have to be ordered in any specific manner and for the problem at hand the Bethe equations \eqref{Bethe2} read 
\beq
\label{Bethe3}
\frac{E_n }{2} \log \( \frac{E_n }{2 \pi  e} \)  -  \sum_{j=1}^N   \Im \log \(  1 -  \frac{e^{- i (E_n \log \q_j - \charphi_j) } } { \q_j^{\sigma} }  \) = (n - \tfrac{3}{2} ) \pi
\eeq
where we have shifted for later convenience $n$ by $-1$.   It is implicit that if there are solutions,  then the $E_n$'s depend on $\sigma$ as well.  As we will explain in the next section, the $E_n$'s coinciding with the Riemann zeros arise in the limit $\sigma \to \half^+$ when, for any $j$, we will take $\charphi_j = 0$.

So far,  the discussion is still quite general,  and there are no convergence issues if $N$ is finite. For our purposes we choose to take $\q_j$ to be the $j$-th prime number,  i.e. $\q_j = \p_j$ where 
$\{ \p_1, \p_2,  \p_3, \ldots \} =  \{ 2, 3, 5, \ldots \}$.  For simplicity let us first consider $\charphi_j = 0$.    Let's mention, en passant, that another interesting choice would be to take  $\q_j$ as any random integer  
between two consecutive primes  since even in this case the quantized energies $E_n$ would equal the Riemann zeros when $\sigma = \half$ \cite{Grosswald} and we will address the study of this case somewhere else.
 Now,  if $\sigma>1$,  the sum over scattering phases in \eqref{Bethe3} converges as $N \to \infty$ 
  and, when  $\pprime_j$ is the $j$-th prime number, the equation becomes
\beq
\label{FLasymptotic} 
\frac{E_n }{2} \log \(\frac{ E_n }{2 \pi  e }  \)  + \arg \, \zeta (\sigma + i E_n )= (n - \tfrac{3}{2} ) \pi
\eeq
where $\zeta(s)$ is the Riemann zeta function, defined both as a Dirichlet series on the integers $n$  and an infinite product on the primes $\p$
\beq
\zeta(s) \,=\,\sum_{n=1}^{\infty} \frac{1}{n^s} \,=\,\prod_{j=1}^{\infty} \frac{1}{1 - \frac{1}{\p_j^s}} \,\,\, ~~~~~(\Re (s) > 1 ), 
\label{zetaRiemann}
\eeq
where we have introduced the complex variable $s = \sigma + i E$.   
As mentioned above,  the quantized energies $E= E_n (\sigma) $ depend on $\sigma$,  however we only display this dependence when necessary.

It is important to remark that $\arg\, \zeta $ in \eqref{FLasymptotic} is the true phase that keeps track of branches, and not $\Arg\, \zeta = \arg\, \zeta ~ {\rm mod} ~ 2 \pi$,  where $\Arg$ is the principle branch with $-\pi < \Arg < \pi$\footnote{The function $\arg \, \zeta (s)$ can be defined by piecewise integration (see Section V).}.  
For $\sigma>\half$,  or for a truncated Euler product at finite $N$,
\beq
\label{argzeta} 
\arg \, \zeta (s) = -\sum_{j=1}^N \, \arg \( 1 - \inv{\p_j^s} \), 
\eeq
it is straightforward to see that each term in the above equation is on the principal branch.    The branch cut is along the negative real axis in the complex $s$ plane.    Thus a change of branch only occurs when the  imaginary part of $1 - \inv{\p^s}$ is zero and its real part negative.      One has 
\beq
\label{ReIm} 
 1 - \inv{\p^s} = \( 1-\frac{\cos (E\log \p )}{\p^\sigma} \) + i \(\frac{\sin (E \log \p )}{\p^\sigma} \).   
 \eeq
 When the imaginary part is zero at $E \log \p = n \pi$,  then the real part equals $1 \pm 1/\p^\sigma$ which is always positive for $\sigma > 0$.  
 Thus 
 \beq
 \label{Argarg}
 \arg \( 1 - \inv{\p^s} \) = \Arg \( 1 - \inv{\p^s} \)~~~~~~~~~~(\Re(s)  > 0).
 \eeq 
Although each term in \eqref{Bethe3} is an $\Arg$ by the nature of the scattering phase,  we know that there will be many cancelations since the average of 
$\arg\, \zeta $ along the critical line is zero (See Section V). 
However the sum of $\Arg$'s can accumulate,  i.e. the sum of $\Arg$'s is not $\Arg$ of the sum such 
that $\arg \, \zeta$ is not always on the principal branch.   It is in fact unbounded on the critical line 
(see  Section V for more  specific remarks).   In summary we will be using 
 \beq
 \label{argsummary}
 \arg \, \zeta (s) = - \sum_\p  \Im \log \( 1 - \inv{\p^s} \).
 \eeq

When $\sigma   > 1$, there is a unique solution to  the equation \eqref{FLasymptotic} for every $n$ since for $E_n$ sufficiently large, i.e. $E_n > \CE_0$ (and it is sufficient to take the low value  $\CE_0 \sim 10$), the left hand side of the equation is a monotonically increasing function of $E_n$. In order to get inside the critical strip to the right of the critical line requires $\half \leq  \sigma  \leq  1$ where the Euler product does not converge,  thus one has to regularize the $N \to \infty$ limit in some manner and the values $E_n (\sigma)$ may depend on such a regularization.  In the following we present three options to deal with this delicate issue.
 
 \subsection{Three options for getting into the critical strip}
 
Although there may be other possibilities, let's here discuss three possible options to deal with the thermodynamic limit of our scattering problem  when  $\half \leq  \sigma  \leq  1$. 
 \begin{itemize}

 \item  The first, and simplest option is to just declare that $\zeta(s)$ in \eqref{FLasymptotic} is the standard analytic continuation presented by Riemann to $\half  \leq  \sigma \leq 1$.  
This is the analog of the regularization  used to handle divergent series present in many quantum mechanics examples, e.g. in the evaluation of the ground state energy of a quantum field theory.

\item The second option is to conveniently modify the original problem by reintroducing non-zero angles $\charphi_j$ in \eqref{Sjs}  based on a  primitive non-principal  Dirichlet character $\character$.
The Dirichlet characters are phases which are roots of unity,  and we make the identification  $ e^{i \charphi_j} = \character (j)$. 
The $\zeta$ function is now replaced by the Dirichlet  $L(s, \character)$ function.     Due to the completely multiplicative property of the characters,  
 $\character (n) \character (m) = \character (nm)$,   the Euler product takes the form
\beq
\label{EPFchi}
L(s, \character) = \sum_{n=1}^\infty  \frac{\character(n)}{n^s}  = \prod_{j =1}^\infty \( 1 - \frac{\character (\p_j )}{\p_j^{s}} \)^{-1}, ~~~~~ \Re(s)  > 1.
\eeq
It is believed that the RH is valid for  this class of functions also, which is  referred to as the General Riemann Hypothesis (GRH).  
Dirichlet $L$-functions of  primitive non-principal Dirichlet characters have no pole at $s =1$ thus it is possible that their Euler product converges to the right of the critical line $\sigma = \half$.  
    In fact it was argued that this is indeed the case due to
a random walk property of the sum $\sum_\p \character(\p)$ arising from the pseudo-randomness of the primes \cite{LMDirichlet}.   The reason for this is that the phases 
$\character(\p)$ over the primes  $\p$ behave as independent, identically distributed random variables.   
For the scattering problem relative to the $L(s, \character )$ functions, one also needs to change the asymptotic dispersion relation \eqref{Dispersion2} to $p = E \log (q E/2\pi e)$ where $q$ is the modulus of $\character$
in order to relate the Bethe ansatz equation to its zeros,  
and the analog of \eqref{FLasymptotic} becomes
 \beq
\label{FLasymptotic2} 
\frac{E_n }{2} \log \(\frac{ q E_n }{2 \pi  e }  \)  + \arg \, L(\sigma + i E_n , \character) = (n - \tfrac{1}{2} ) \pi.
\eeq
Truncating the infinite Euler product to the integer $N$, the corresponding truncated expression $\arg_N \, L(\sigma + i t)$ 
for the argument of the Dirichlet function is given by 
\beq
\arg_N \,L(\sigma,t) \,=\,
 \label{argsummary}
 - \sum_\p^N  \Im \log \( 1 -\frac{\character (\p)}{\p^s} \).
 \eeq
 Expanding the logarithm, we have 
 \beq 
\arg_N \,L(\sigma,t) \,\simeq  \Im \left(\sum_p^N \frac{\character (\p)}{\p^s}\) 
\eeq
With respect the Riemann $\zeta$-function, for the properties mentioned above of the non-principal characters computed on the primes $\character (\p)$ we expect the truncated series for the argument of the Dirichlet function has a better convergent behaviour, a fact that is indeed well confirmed by comparing the truncated expression with the actual values 
of the argument of the Dirichlet function (see for instance Figure \ref{ArgDirichlet55} for a very good agreement of $\arg \, L$ for the mod $q=5$ Dirichlet character 
indicated below in eq.\,\eqref{chis} and its truncated expression $\arg_N \, L(\sigma + i t)$ which makes use of just  $N=100$.)

\begin{figure}[t]
\centering\includegraphics[width=.4\textwidth]{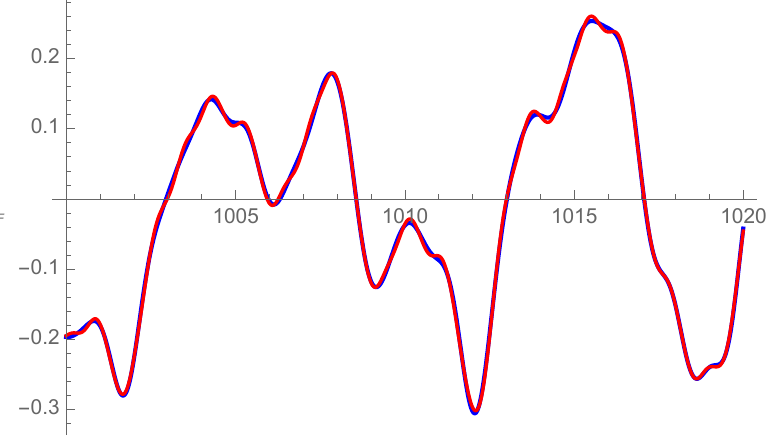}
\caption{A plot of $\arg \, L(s,\character)$ for the Dirichlet $L$-function based on the character $\character$  given in \eqref{chis} for $s=0.55+ i t$ as a function of $t$ around $t=1000$.   
The red curve is computed from the analytic continuation into the critical strip, the blue one is computed from the Euler product using just $N=100$ primes. As evident from the figure, the two curves are hardly distinguishable.   
}
 \label{ArgDirichlet55}
\end{figure} 

 \item
Thirdly, for the original case related to the Riemann $\zeta$-function, one can simply consider the scattering problem for a {\em finite} number $N$ of impurities,  $N< N_c$,   so there are no convergence issues. As a matter of fact, it is known that a truncated Euler product for finite $N_c$  can be a good approximation to $\zeta$ for $\sigma> \half$  if the truncation is well chosen.  Of course this third option is the least desirable since, in principle, the values of $E_n$ could be sensitive to 
$N_c$.    It was argued in \cite{FL2} that, if for a given $s$, one truncates to $N_c  \sim E^2 $ primes,  then this is a good approximation to $\zeta$ to the right of the critical line $\sigma > \half$.    Of  course if one goes beyond this optimal truncation, as it happens in any other asymptotical expression, then the truncated product will eventually start to diverge.    However for very high $E_n$,  the number of primes is huge so that a finite truncation can be a good approximation.  
 In other words as $E\to \infty$,   the truncation number $N_c \to \infty$.   
 Henceforth,  finite Euler products inside the critical strip for $s= \sigma + i E$ are implicitly truncated to $N< N_c = [E^2]$. 
 See, for instance, Figure \ref{ArgZeta55} for illustration of the validity of the truncated Euler product in a certain interval. 
 Let $E_{n;N}$ denote the solution to \eqref{Bethe3} using $N$ primes versus $E_n$ computed from \eqref{FLasymptotic},  both for $\sigma = \half$.       
 Then in \cite{ALZeta} the error was estimated as 
 \beq
 \label{error}
 \frac{E_n - E_{n;N}}{2 \pi/ \log n} \approx \inv{\pi \sqrt{\log N} } \cos \( E_n \log \p_N \)
 \eeq  
where $2 \pi/  \log n$ is the average spacing between zeros.     Note that as $N \to \infty$ the error goes to zero. It is also worth mentioning that, 
assuming the Riemann hypothesis,   Gonek provided a much more detailed analysis which expresses the error in terms of zeros of $\zeta$ on the critical line \cite{Gonek}.

  \begin{figure}[b]
\centering\includegraphics[width=.4\textwidth]{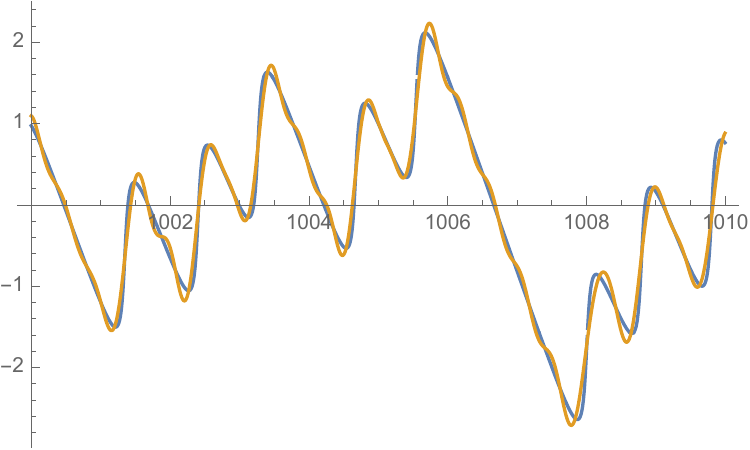}
\caption{A plot of $\arg \, \zeta (s) $  for $s=0.55+ i t$ as a function of $t$ around $t=1000$.   
The blue curve is computed from the analytic continuation into the critical strip,  the yellow one is computed from the Euler product  using $N=100,000$ primes.    
}
 \label{ArgZeta55}
\end{figure}

\end{itemize}

Summarizing, the significant difference between Dirichlet $L$-functions based on principal characters,  such as $\zeta (s)$ itself,   and others based on non-principal characters is 
merely due to the pole at $s=1$ for the principal case, and this pole is responsible for the need to truncate the Euler product  in the principal 
case\footnote{As shown in \cite{MLRW}, a way to approach the Riemann Hypothesis for the $\zeta$-function which avoids truncation consists to consider the 
M\"obius function $\mu(s) = 1/\zeta(s)$ which does not the pole at $s=1$. } \cite{LMDirichlet}.

Let us also comment that 
there are an infinite number of potentially interesting scattering problems based on the generalized zeta functions defined by the infinite product of our scattering matrices, but for simplicity here we will mainly consider in detail the three options presented above. In particular, in the first option, as $\sigma \to \half^+$,   the quantized energies  $E_n (\sigma)$  satisfying \eqref{FLasymptotic} asymptotically approach the Riemann zeros on the critical line $\sigma = \half$  as $n \to \infty$.  The equation (\ref{FLasymptotic}) was first proposed in \cite{Electrostatic} and, as a matter of fact, is not very asymptotic at all. For the lowest zero at $n=1$,   with  $\sigma = \half + \delta$, and    $\delta =  0.0001$,  one finds $E_1 = 14.1347$,   which is correct to 6 digits.   By systematically reducing $\delta$ one can calculate 
the Riemann zeros to great accuracy from the exact equation described in the next section,    from  hundreds to even thousands of digits for even relatively low  zeros \cite{FrancaLeClair}.       In this article we limit the numerics to zeros around the $100$-th for illustration,    but very similar results apply to much higher zeros.    
From \eqref{FLasymptotic} we obtain
\beq
\label{En100} 
\lim_{\sigma \to \half^+}  \{ E_{100}, E_{101}, \ldots , E_{104} \}  = \{ 236.524, 237.769, 239.555, 241.049, 242.823\},
\eeq
which are identical  to the true Riemann zeros to the number of digits shown\footnote{In this range of $E$ one can easily check that $\Arg = \arg$ for $\zeta ( \half + i E)$ at the zeros $E_n$.    Here we have also taken $\delta = 0.0001$ however one can take it to be much smaller to improve the accuracy almost indefinitely \cite{FrancaLeClair}.}.
If one adopts the third prescription of a finite number of impurities, one  still obtains good results.  
For only $N=1000$ impurities, for example, one finds
\beq
\label{En100FiniteN3} 
\lim_{\sigma \to \half^+} \{ E_{100}, E_{101}, \ldots , E_{104} \}  = \{ 236.521, 237.777, 238.139, 241.057, 242.812 \} , ~~~~~{\rm for~} N=1000 ~ {\rm impurities}. 
\eeq

 \subsection{Dependence of $E_n$ on  $\sigma$}

 It should be emphasized that the $E_n (\sigma )$ do not correspond to Riemann zeros unless $\sigma = \half + \delta$ where $\delta = 0^+$.  
 However the dependence on $\sigma$ is interesting to consider.  
   By differentiating \eqref{FLasymptotic} with respect to $\sigma$ one easily obtains    
 \beq
 \label{DEn}
 \frac{\d E_n (\sigma)}{\d \sigma} = - \frac{ \Im \( \zeta'(s)/\zeta(s) \) }
 {\Re \( \zeta'(s)/\zeta(s) \) + \log (E_n (\sigma)/2 \pi )/2 }, 
 ~~~~~ \zeta' (s) = \d_s \zeta(s) , ~~~~~~ s = \sigma + i E_n (\sigma),
 \eeq
 where 
 \beq
 \label{zpz} 
 \frac{\zeta' (s)}{\zeta(s)} = - \sum_{\p} \frac{\log \p}{\p^s -1 }, ~~~~~~~~(\Re (s) > 1).
 \eeq

\begin{figure}[t]
\centering\includegraphics[width=.4\textwidth]{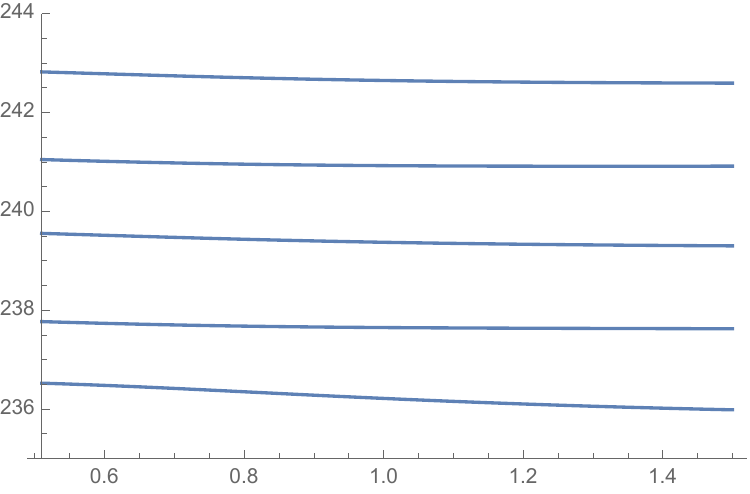}
\caption{A plot of $E_n (\sigma)$ for $n = 100,  101, \ldots, 104$ as a function of $\sigma$  to the right of the critical line.    
When $\sigma = 1/2$ the $E_n$ correspond to the $n$-th  Riemann zero on the critical line.}
 \label{EnSigma1}
\end{figure}

\begin{figure}[b]
\centering\includegraphics[width=.4\textwidth]{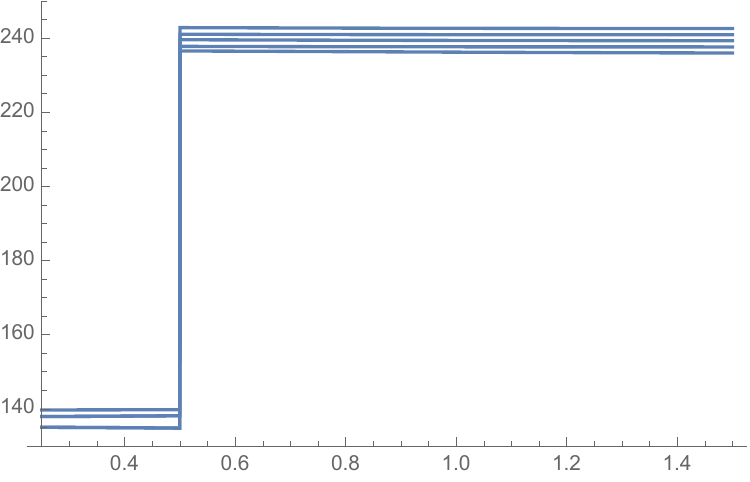}
\caption{A plot of $E_n (\sigma)$ for $n = 100,  101, \ldots, 104$ as a function of $\sigma$ across the critical line.}
 \label{EnSigma2}
\end{figure}

\begin{figure}[t]
\centering\includegraphics[width=.4\textwidth]{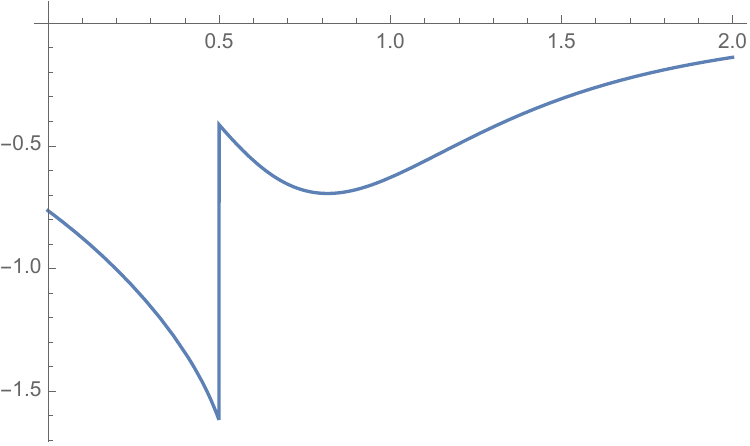}
\caption{A plot of $\d_\sigma E_{100} (\sigma)$ as a function of $\sigma$.}
 \label{EnSigma3}
\end{figure}

 Interestingly,   the $E_n (\sigma)$ do not vary much for $\half< \sigma < 2$,   as can be seen in Figure \ref{EnSigma1}.    
 In fact for higher $n$ the curves become flatter,  which can be attributed to the $\log (E_n (\sigma)/2 \pi )$ in \eqref{DEn} which goes to infinity as $n\to \infty$.   
 This is an important fact since for $\sigma > 1$ the Euler product absolutely converges and there is a unique solution to \eqref{FLasymptotic} in this regime of $\sigma$ and $E_n (\sigma)$ is smoothly deformed to the Riemann zeros as $\sigma \to \half$.      However this feature abruptly changes to the left of $\sigma =\half $ where $\d_\sigma E_n (\sigma)$ diverges,    as can be seen 
 in Figures  \ref{EnSigma2}, \ref{EnSigma3}.         This behavior is  obviously indicative of zeros on the critical line.    
 
\bigskip

\section{Scattering problem for the exact  Riemann zeros}

In this section we refine the model of the previous section in such a way to give the {\it exact} Riemann zeros on the critical line.   
This will only require a small correction to  the dispersion relation, a correction which vanishes as $E \to \infty$.
For simplicity we present the results for $\zeta (s)$,  even though similar exact equations for the zeros of Dirichlet $L(s,\character)$ can be easily formulated 
\cite{FrancaLeClair}.   
 Following standard conventions in analytic number theory,  let  us define a complex variable $s= \sigma + i t$ where, based on the notation above, 
  $t= E$.   We consider zeros {\it on the critical line},  which are known to be infinite in number \cite{Hardy}. Denote the $n$-th zero 
 on the upper critical line as
\beq
\label{rhozero}
\rho_n = \half + i t_n, ~~~~~n=1,2,3, \ldots
\eeq
where $t_1 = 14.1347..$ is the first zero, and so forth. 
Labeling them this way, we define below an impurity scattering problem where $ E_n (\sigma)$ of the previous section become the exact $t_n$ as $\sigma \to \half^+$.          

Define a completed $\zeta$ function as follows\footnote{As we see later in Section VI, Riemann defined an entire function  which is the above $\chi  (s) $ multiplied by $s(s-1)/2$  in order to cancel the simple pole at $s=1$,   however this is not necessary here for our discussion. }:
\beq
\label{chidef}
\chi (s) = \pi^{-s/2} \Gamma(s/2) \zeta (s),  
\eeq
which satisfies the non-trivial functional equation 
\beq
\label{func}
\xi(1-s) = \xi(s).
\eeq
Let us now write $\xi (s)$  in terms of a positive, real modulus $|\xi(s)|$  and argument $\theta$:
$\xi(s) = |\xi (s) |  \, e^{i \theta (\sigma,t)}$,  i.e. 
\beq
\label{argchi}
\theta (\sigma,t) = \arg\, \xi (\sigma + it).
\eeq
As discussed in the last section,  
  it is important that $\theta$  is the true $\arg$,  not $\Arg = \arg ~ {\rm mod} ~ 2 \pi$,   where $\Arg$ is the principle branch with $-\pi < \Arg < \pi$. 
Obviously 
\beq
\label{thetasig}
\theta (\sigma, t)  = \thetaRS(\sigma,t) + \arg \, \zeta (\sigma+ i t) 
\eeq
where
\beq
\label{varthetadef}
\thetaRS(\sigma, t) \equiv \arg \, \Gamma \( \half (\sigma + i t) \) - \tfrac{t}{2} \log \pi.
\eeq
On the critical line $\thetaRS (\half, t)$ is smooth and  commonly referred to as the Riemann-Siegel $\thetaRS$ function.  
Below,  if it is implicit that we are on the critical line we will simply write $\thetaRS (t) \equiv \thetaRS (\half, t)$.  

On the critical line,  $\xi (s)$ is real due to the functional equation and it is zero for $s=\half + i t_n$ (see Figure \ref{ChiRegAlternating} for the plot of $\xi(s)$ along the critical line). Thus moving up the critical line, $\theta(s)$ must jump by $\pi$ at each {\it simple} zero where it changes sign.   We will call this the {\em vertical approach}.    
 Of course we can also approach a zero from other directions and again relate
$t_n$ to a specific angle.  We will consider both vertical and horizontal approaches, defined more precisely as follows \cite{FrancaLeClair}.

\bigskip
\noindent
{\bf Vertical approach}

Approaching a zero along the critical line from above:
\beq
\label{Vertn}
\lim_{\epsilon \to 0^+ } \theta (\half, t_n + \epsilon ) = (n-1)\pi 
\eeq
It's important to note that the non-zero $\epsilon$ in \eqref{Vertn} is absolutely necessary:  if $\epsilon=0$ the equation is not well defined 
since $\arg \, \zeta (\half + i t_n )$ is not defined unless one specifies a direction of approach to the zero.

\bigskip
\noindent
{\bf Horizontal approach}

Approaching a zero along the horizontal direction from the right of the critical line:
\beq
\label{Horzn}
\lim_{\delta \to 0^+ } \theta (\half + \delta, t_n ) = (n-\tfrac{3}{2})\pi .
\eeq
This is just a $90^\circ$ rotation of equation \eqref{Vertn}  thus we sent  $n\to n - \half$.  
The advantage of this horizontal approach  is that there are better convergence properties to the right of the line with $\delta >0$. 

\vspace{3mm}
Henceforth we will consider this latter approach for its better convergence properties and the corresponding equations 
can be written as 
 \beq
 \label{FLeqn}
 \thetaRS (t_n) +  \lim_{\delta \to 0^+}  \arg \, \zeta(\half + \delta + i t_n )  = (n - \tfrac{3}{2} ) \pi \,\,\,.
 \eeq
 Two remarks are now in order. First, notice the strict similarity of these equations with the Bethe Ansatz equations (\ref{FLasymptotic}), the only difference is in the first term which refers  to the momentum of the particle as expressed in terms of the energy (here denoted as $t_n$). Hence, assuming our particle has an  asymptotic dispersion relation given by 
 \beq
 \label{DispersionRS}
   p(E)  = 2 \thetaRS(E) = E \, \log \( \tfrac{E}{2 \pi e}\) - \tfrac\pi4 + \CO(1/t) ,
  \eeq
   (see \eqref{thetaStirling} based on the Stirling approximation),  
  we can interpret the equations above for the exact zeros of the Riemann $\zeta$-function as coming from a Bethe ansatz approach. The second remark concerns the completeness of these Bethe ansatz equations, namely it is important to notice that, if for any integer $n$ there is a unique solution of the equations  (\ref{FLeqn}), then the Riemann Hypothesis is true and all zeros are simple. The reason why this statement is correct goes as follows \cite{FrancaLeClair}. Let $N(T)$ denote the number of zeros in the entire critical strip $0< \sigma < 1$ up to height $T$.  Then if $T$ is not a zero $t_n$,  it is known from the argument principle that 
 \beq
 \label{NofT}
 N(T) =  \thetaRS (T)/\pi   + \CS (T)  + 1 
 \eeq
 where $\CS(T)$ is defined in \eqref{SofTFirst}. Then the  solutions to the equation \eqref{FLeqn}  saturate the counting formula $N(T)$.  The shift by $1$ above is due to the simple pole at $s=1$.

\bigskip

One can  check that for all known Riemann zeros, which is quite a large collection,   the equations (\ref{Vertn}) and (\ref{Horzn}) above are exactly satisfied and they 
can in fact be used to calculate  Riemann zeros to high accuracy \cite{FrancaLeClair}.  
Ignoring the $\arg \,\zeta$ term,  for eq.\,(\ref{Horzn}) we have $t_n \approx \ttilde_n$ where
\beq
\label{tminusn}
\thetaRS (\ttilde_n ) = (n-\tfrac{3}{2})\pi .
\eeq
The solutions of these equations can be referred to as {\it anti}-Gram points, namely the values of $t$ where the real part of $\zeta(\half +i  t)$ is zero,  but the imaginary part is non-zero\footnote{The usual well-known Gram points are the opposite,  i.e points where the imaginary part is zero but the real part is non-zero.    They  satisfy $\thetaRS (t) = (n-1) \pi$ and are thus more appropriate to the vertical apprroach
based on \eqref{Vertn}.}.  
One expects these points to be closer to the actual zeros  than the Gram points since it is known that the real part of $\zeta ( \half + i t )$ is  nearly always  positive.   
For large $t$
 one can use the Stirling approximation for $\arg  \, \Gamma = \Im \log \Gamma$ to obtain on the critical line
\beq
\label{thetaStirling}
\thetaRS (t ) = \frac{t}{2} \log \( \frac{t}{2 \pi e} \) - \frac{\pi}{8} + \inv{48 t} + O(1/t^3).
\eeq
For large $t$,  the $t\log t$ term strongly dominates.   

As for $p(E)$ in eq.\,(\ref{Dispersion2}), $\thetaRS (E)$ is monotonic and positive in the infinite interval $(E_*,\infty)$ (the only interval we are interested in), where 
$E_* \simeq 17.8456$,  and in this interval it is invertible and therefore asymptotically eq.\eqref{Eofp} is valid.  
For large $n$ the solution to the equation (\ref{tminusn}) above is approximately 
\beq
\label{tnLambert}
\ttilde _n  \approx  \frac{2 \pi (n-\tfrac{11}{8})}{W\( (n- \tfrac{11}{8})/e \)} \,\,\,.
\eeq

In this limit of large $n$,  the   solution to \eqref{FLasymptotic}  
is approximately a solution to the exact equation  \eqref{Horzn}.
Again the non-zero $\delta$ in \eqref{Horzn} is absolutely necessary,  since if $\delta =0$, 
 $\arg \, \zeta (\half + i t )$ is not defined  at a zero $t= t_n$ unless a direction of approach is specified.

It is interesting to study the behavior of a fixed $t_n = E_n(\sigma = \half^+)$ as one increases the number of impurities,  keeping in mind the optimal truncation that the number of impurities 
$N  \lesssim N_c = [t^2]$.
 Focussing again on $E_{100}$ we present results in the Table \ref{100th}.  There are several important remarks to make.   
 The approximation $\ttilde_n$ based on the Lambert $W$ function is smooth,   and usually gets the first $n$ digits,  namely the integer part,  of $t_n$ correct,   but  has no interesting statistics.  
 For instance $\ttilde_{100} = 235.987$.   
 The random matrix statistics of the Montgomery/Odlyzko conjecture \cite{Montgomery,Odlyzko}
 obviously come from the fluctuations in $\Soft (t)$ 
 \beq
 \label{SofTFirst}
 \CS (t) =\lim_{\delta \to 0^+}  \inv\pi  \arg \, \zeta (\half + \delta + i t ),
 \eeq
  as is evident from Table \ref{100th}.   
 These fluctuations are due to the pseudo-randomness of the primes.  These statistics were reproduced for solutions of  the asymptotic equation  of the last section  and the exact solutions of \eqref{Horzn} in \cite{FrancaLeClair}.

\begin{table}
\begin{center}
\begin{tabular}{|c|c|}
\hline\hline
Number of impurities $N$     &   $E_{100} = t_{100}$     \\
\hline\hline 
$1$       &  $236.386$ \\
$10$     &  $236.521$  \\
$100$   &   $236.512$  \\
$200$    &  $237.336$   \\
$300$    &  $236.525$  \\
$400$    &  $236.526$ \\
$500$    &   $236.511$ \\
$600$   &  $236.525$  \\
 $700$   & $236.532$ \\
 $800$    &  $236.530$   \\
 $900$   &    $236.528$   \\
 $1000$  &  $236.521$  \\
 $10,000$  & $236.524$  \\
\hline\hline 
\end{tabular}
\end{center}
\caption{The $100$-th energy $E_{100}$  computed with an  increasing number of impurities $N$.    The actual Riemann zero is $t_{100}= 236.524$. }
\label{100th}
\end{table}

 \vfill\eject

\section{Pertinent known properties of \large{$\Soft (t)$}}

As evident from our previous considerations, a crucial object in our Bethe Ansatz approach is the function 
$\Soft (t)$ defined above in \eqref{SofTFirst}.    It is then important to remind the main properties of such a function 
(for a review see \cite{Karatsuba}).  Although we do not display the $\delta \to 0^+$ limit in \eqref{SofTFirst},   it is implicit
throughout this section.    This is important since without this $\delta$,    it is known that $\Soft (t)$ jumps discontinuously by $1$ at each zero 
{\it on the line.}  

\bigskip

(i)  
A classical result of Bohr and Landau \cite{BL} states that, when $t$ increases, $\Soft (t)$ has infinitely many sign changes and its average 
is zero \cite{Edwards,Titchmarsh}.

\bigskip

(ii)  $\CS (t)$ is unbounded.   Von Mangoldt first showed that $\CS(t) = O (\log t )$.    Backlund computed specific bounds in 1918.       
The most recent bound we could find is due to Trudgian \cite{Trudgian}  which is only a modest improvement of Backlund's bound:
\beq
\label{Trudgian} 
|\CS(t) | \leq 0.1013 \log t .
\eeq
This result does not assume the Riemann Hypothesis (RH).   Assuming instead the RH, it was shown that $\CS (t) = O \( \tfrac{\log t}{\log \log t} \)$ 
\cite{GonekGoldstone}.
In fact the largest value of $\CS (t)$ so far observed in computations around the $n=10^{30}$-th zero is roughly  $3.3455$ \cite{Ghaith}.

\bigskip

(iii)  A celebrated theorem of Selberg \cite{Selberg} states that $\CS(t)$ over a large interval $0 < t < T$ satisfies a normal distribution with zero mean and variance
\beq
\label{Selberg} 
\overline{\CS(t)^2}   \equiv \inv{T} \int_0^T  \CS (t)^2 \, dt = \inv{2 \pi^2}  \log \log T + O( \sqrt{\log \log T} ).
\eeq
This is a very interesting property for our purposes since in order to derive this result,   for the $2k$-th moment of $\CS(t)$ one needs to  {\it truncate} the Euler product to
primes $\p<x$ where $x< T^{1/k}$. Thus this means that a finite $N$ number of impurities can capture important properties of the Riemann zeros.    For the second moment one should truncate at $\p < T$,   thus for large $T$ the scattering problem still converges for a  very large number $N$ of impurities.      

 \bigskip
 
All properties above have clear implications for this article.     First, the $\arg \, \zeta$ term in \eqref{FLeqn} being $O(\log t)$ is strongly subdominant compared to the monotonic 
$t \log t $ term coming from $\thetaRS (t)$. Hence, for large enough $t$, one expects that this last term will dominate the left hand side of the equation (\ref{FLeqn}) 
and therefore that there should be a solution of this equation for every $n$.   
   Secondly,    Selberg's central limit theorem shows that a finite number $N$ of impurities is a meaningful approximation to the Euler product if one truncates it properly, ensuring continuity of the function in the left hand side of the Bethe ansatz equation (\ref{Bethe2}). For more rigorous work on the validity of truncated Euler products assuming the RH, we refer again  to Gonek's work \cite{Gonek}.

 \section{A definition and some analysis}

An important question is the following.  If the RH is false,   then what is the signature of zeros off the line in the real  function $\chi (\half + i t )$, 
if any?    
For  this purpose,  it
 will be useful in the sequel to define a ``regular alternating function":

\bigskip

{\bf Definition:} ~~  We define a   {\it regular alternating function} (RAF)  as  a real function which has only one minimum or maximum between its zeros. 

\bigskip

\noindent
The simplest known examples are the sine and cosine functions or the Bessel functions of first kind $J_n(x)$. 
This definition will also be  useful below where we consider the Davenport-Heilbronn counterexample.

\def\rhozero{\rho_{\bullet}}
\def\rhozeroConj{\bar{\rhozero}}
\def\tzero{t_\bullet}

\bigskip

\def\fhat{\hat{f}}
\def\chitilde{\tilde{\chi}}

\begin{figure}[t]
\centering\includegraphics[width=.4\textwidth]{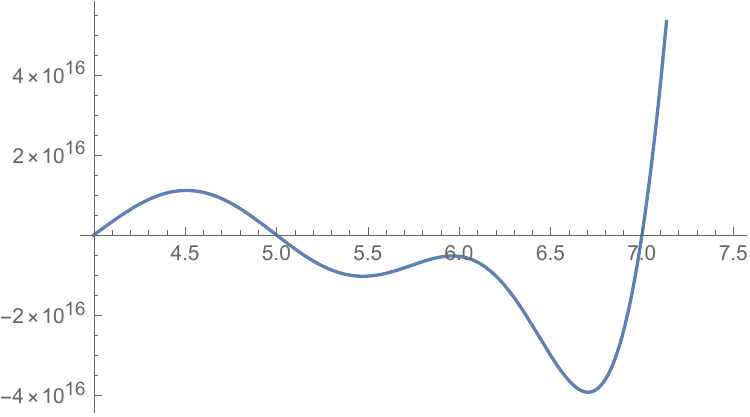}
\caption{A plot of $\fhat (\half + i t )$ defined in \eqref{chiMulti} as a function of $t$ in a region where there are zeros off the line.     The function $\fhat (s)$ has zeros on the line 
at $\tzero = 1, 2,3, 4,5, 7, 8,9,10$ and a double zero at $t=6$:
$\rho_* = \tfrac34 + 6i$.}
 \label{ChiWithDouble}
\end{figure}

We now argue that if $\chi (\half + it)$ is a regular alternating function, then the RH is true.  
Let us first motivate this with a simplistic  example.    
First define a function  $f(s)$ which has a 2 simple zeros on the line at $\rhozero = \half \pm  i \tzero$:
\beq
\label{chisingle}
f_{\rm single} (s,  \tzero) =( s - (\half + i \tzero )) (1 - s - (\half+ i \tzero)).
\eeq
Next define an  $f(s)$ with 4 simple zeros off the line $\Re(\rho_*) \neq \half$,  at $s= \rho_*, 1 -\rho_*$ and their two complex conjugates \:
\beq
\label{chidouble}
f_{\rm double} (s, \rho_*) = 
(s - \rho_*)  (1 - s - \rho_*) (s - \overline\rho_*) ( 
  1 - s - \overline\rho_*).
  \eeq
  Both the above $f(s)$ are constructed to satisfy the functional equation $f(s) = f(1-s)$ and are thus real on the critical line $\Re (s) = \half$.     
  From these building blocks we can construct a function with both zeros on and off the line.  
  For example, the following  $\fhat (s)$, for  $t>0$,  has the first  five zeros on the line at $s= \half + i n$, $n=1,2, \ldots, 5$ 
  a pair of zeros off the line at $ s= \tfrac34 \pm 6i$,  followed by $4$ zeros on the line at 
   $s= \half + i n$, $n=7,8, 9, 10$:
  \beq
  \label{chiMulti}
  \fhat (s) =
  \(  \prod_{n=1}^5  f_{\rm single} (s, n) \) \cdot f_{\rm double} (s, \tfrac34 + 6i) \cdot \(  \prod_{n=7}^{10}  f_{\rm single} (s, n) \).
  \eeq
 One sees from Figure \ref{ChiWithDouble} that  the regular alternating property of $\fhat (\half + i t)$
 is spoiled  precisely at the ordinate of the zeros off the line.    The function is headed  for a zero at $s= \half +  6i$ but then turns around before reaching it.  

\begin{figure}[t]
\centering\includegraphics[width=.9\textwidth]{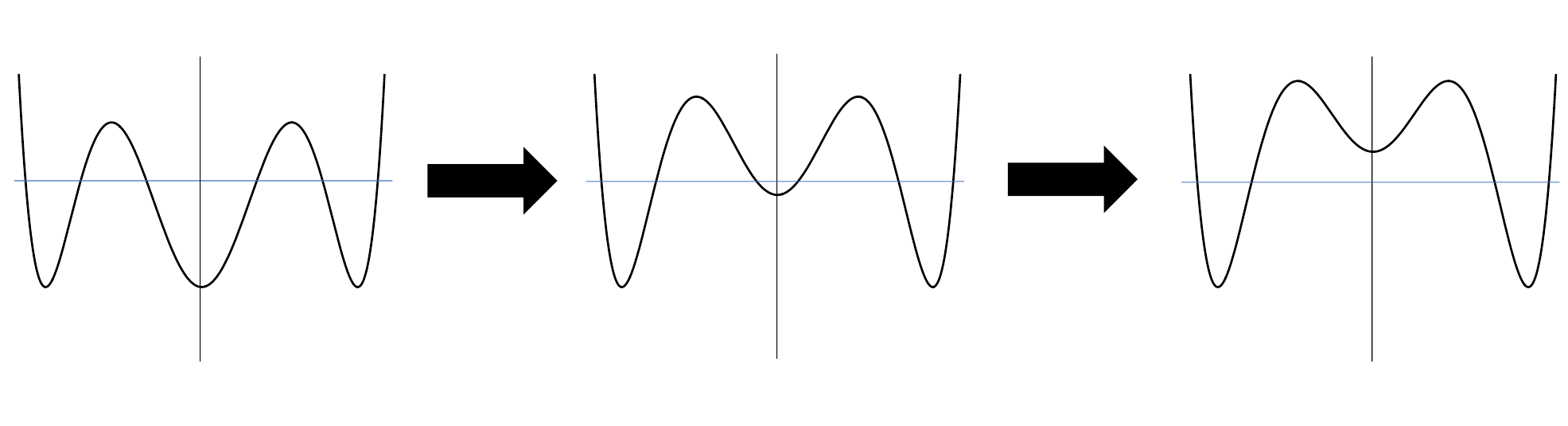}
\caption{A continuous evolution of a regular alternating function into  one that is not (from left to right), losing two zeros on the way.}
 \label{zerooooo}
\end{figure}

More generally,   the easiest way to convert a regular alternating function into one that is not is to decrease one of its maxima or to increase one of its minima, as shown in Figure \ref{zerooooo} with an example in which the minimum of a regular alternating function is continuously pushed up. When the minimum crosses the horizontal $t$-axis, {\em two} zeros are missed at once, becoming complex.

\begin{figure}[b]
\centering\includegraphics[width=.35\textwidth]{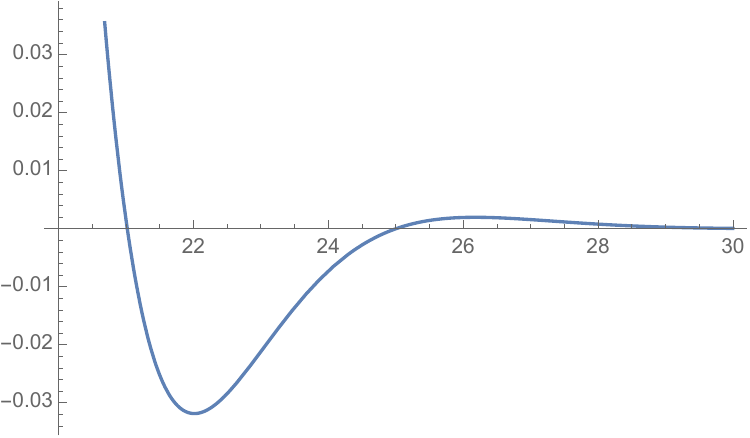}
\caption{A plot of $\chi (\half + i t )$ as a function of $t$ in the region covering zeros $t_n$ for $n=2,3,4$.  (Magnified by $10^{6}$.)}
 \label{ChiRegAlternating}
\end{figure}

For some  modest numerical evidence that $\chi (\half + it )$  is a RAF see Figure \ref{ChiRegAlternating} 
and  one could  argue that  the exact $\chi(\half  + i t )$ is a RAF as follows.    Let us write  
 \beq
 \label{regalt1}
 \chi ( \half + \delta + i t) = |\chi ( \half + \delta + i t)| \,  \exp \( i \theta(\half + \delta + i t ) \) ,
 \eeq
and noticing that $\chi ( \half + i t)$ is real as $\delta \to 0$, we have  
 \beq
 \label{regalt2} 
 \lim_{\delta \to 0^+} \chi(\half + \delta + it ) = \lim_{\delta \to 0^+} |\chi ( \half + \delta + i t)|  \, \cos \( \theta(\half + \delta + i t) \).
 \eeq
So, if $\lim_{\delta \to 0^+}  \theta(\half + \delta + i t) $ is well-defined and monotonic then $\lim_{\delta \to 0^+} \chi(\half + \delta + it )$ is a 
RAF 
since $\cos (x)$ is a regular alternating function of $x$.  There are indications that this is indeed the case since, as we have argued previously, the Euler product renders  $\lim_{\delta \to 0^+}  \theta(\half + \delta + i t) $ well defined and it is monotonic because of its relation to the counting function $N(T)$ \eqref{NofT}. We will show how this property 
can be violated in the next section for a counter example.

\begin{figure}[t]
\centering\includegraphics[width=.35\textwidth]{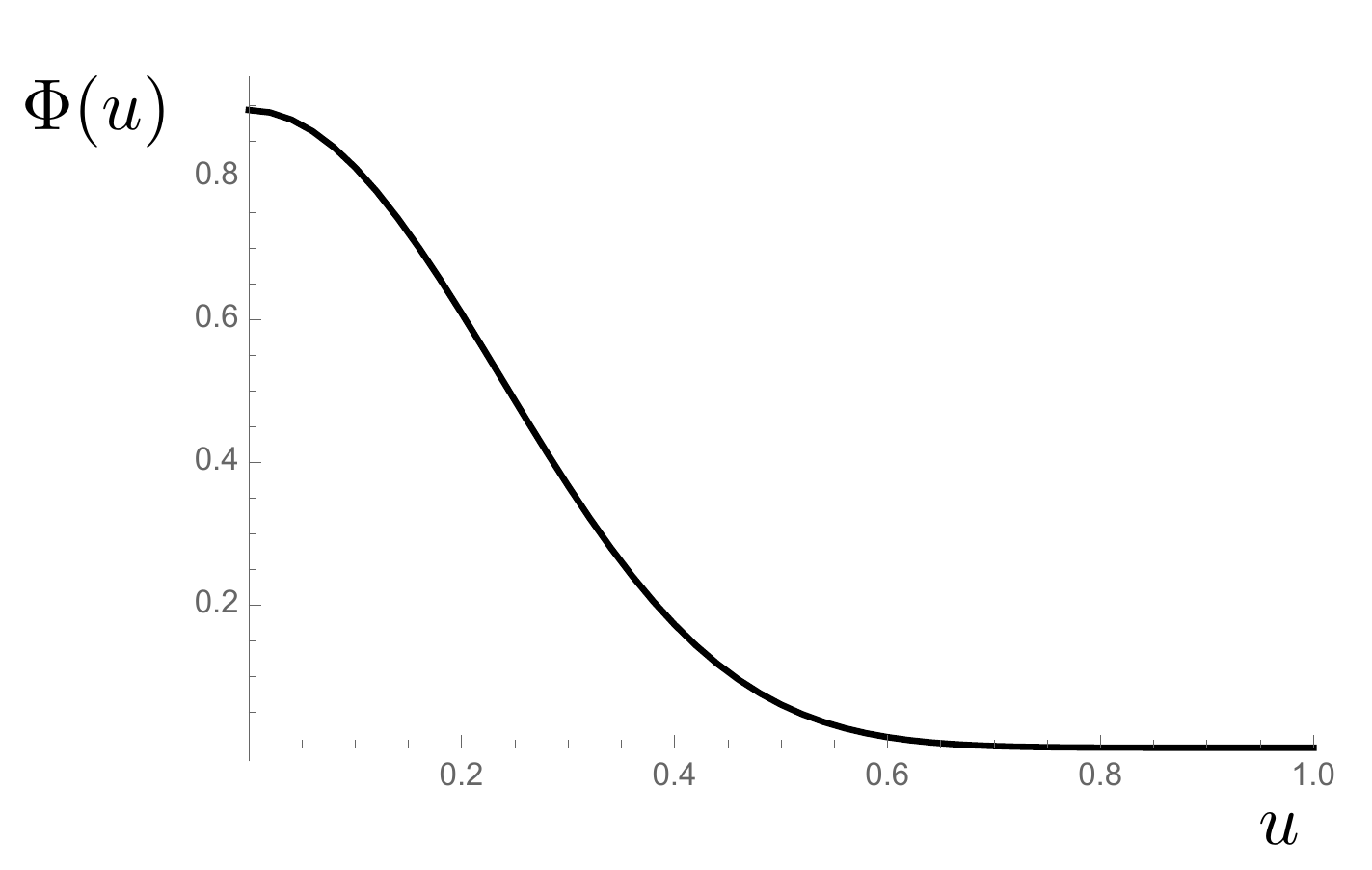}
\caption{A plot of $\Phi(u)$ as a function of $u$  which shows the rapidly decreasing behaviour of this function.}
 \label{Phiuu}
\end{figure}

In light of the above remarks,  let us now return to the  $\zeta (s)$  function and provide some supporting analysis.  
   The regular alternating property of a simple RAF  such as $\cos (t)$ is 
easily spoiled by a simple shift $\cos (t) \to \cos (t) + c$ where $c>1$ is a constant,  which alters the property that its average is zero.  Thus let us first show 
that the average of $\chi (\half  + i t)$ equals zero.   
One can remove the pole in $\chi (s)$ by multiplying by $s (s-1)/2$ which preserves the functional equation.    
\beq 
\chitilde (s) \, \equiv  \tfrac{1}{2} s \,(s-1)  \, \chi (s) = \tfrac{1}{2} s \,(s-1) \, \pi^{-s/2}  \, \Gamma\left(\tfrac{s}{2}\right)  \,\zeta(s) .
\eeq
This  completed Riemann zeta  function is real once specialized on the critical line $s = \half + i t$ (hereafter simply denoted $\chitilde (t)$) and admits the 
remarkable Fourier transform which was stated in Riemann's original paper (see also Chapter 10 in \cite{Titchmarsh}): 
\beq 
\chitilde(t) =\, 2 \int_0^{\infty} \Phi(u) \cos ut \, du \,\,\,,
\label{FFTT}
\eeq
where the spectral function $\Phi(u)$ is given by 
\beq
\label{Phi} 
\Phi(u) \,=\, 2 \pi  e^{5u/2} \,\sum_{n=1}^\infty n^2 \left(2 n^2 \pi e^{2u} - 3 \right) \, e^{-n^2 \pi e^{2 u}}.
\eeq    
The latter can be expressed in terms of  one of the Jacobi   $\Theta$ functions and its derivatives  using 
$(\Theta (x) -1)/2 = \sum_{n=1}^\infty e^{-n^2 \pi x}$.   
Indeed,   the functional equation $\chitilde  (s) = \chitilde (1-s)$ follows from the fact that $\Theta$ transforms as a weight $1/2$ modular form:
$\Theta(1/x) = \sqrt{x} \, \Theta (x) $.    
Three important properties of $\Phi(u)$ are: ~(1) this function is defined by an incredibly fast convergent series;  ~(2) $\Phi(u)$ is a positive function for all $ u > 0$; ~(3) $\Phi(u)$ 
decreases very rapidly with  increasing $u$, so rapidly that its only appreciable finite values are up to  $u \sim 1$ (see Figure \ref{Phiuu}). Hardy made use of such a 
Fourier transform and other related results to prove that  $\tilde \chi(t)$ changes sign infinitely many times, i.e. there are infinitely many zeros on the critical line \cite{Hardy}. 
Using the Fourier expression of $\tilde\chi(t)$ given above,  one can easily show  that the average of  $\chitilde(t)$ is  zero.  
Indeed the average of $\tilde\chi(t)$ on an interval $(0, T)$ is 
\beq 
\langle \tilde \chi\rangle_T \,\equiv \frac{1}{T} \int_0^T \tilde \chi(t) \,dt   
 \,= \frac{2}{T}  \,  \int_0^{\infty}  \frac{du}{u}  \,   \Phi(u) \,{\sin (uT)}
 \label{meeen}. 
\eeq
Now,   one has 
\beq
\label{Phimax}
\Phi (u) \leq \Phi(0) = 0.893394......, ~~~~~~~~~~ {\rm for} ~ u>0.
\eeq
Thus 
\beq
\langle \tilde \chi\rangle_T <  \frac{2 \Phi(0)}{T} \,  \int_0^\infty \frac{du}{u} \sin (uT) \,=\, \frac{\pi \Phi(0)}{T}
 ~~~~~~\Longrightarrow ~~~
\lim_{T \to 0} \langle \tilde \chi\rangle_T =0 .
\eeq

\def\Ei{{\rm Ei}}

\begin{figure}[t]
\centering\includegraphics[width=.35\textwidth]{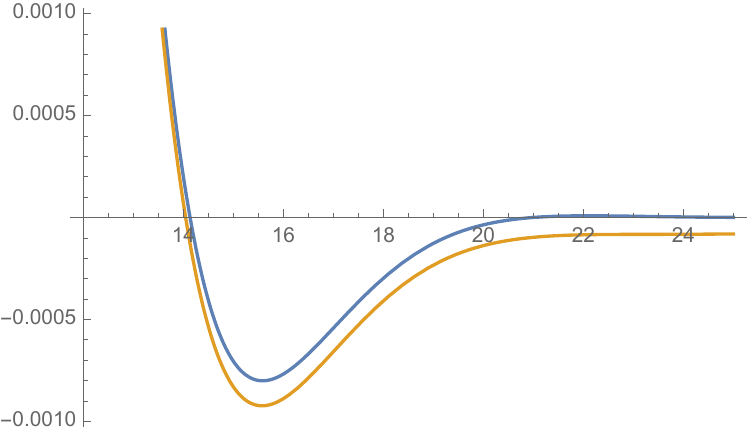}
\caption{A plot of the exact  $\chitilde (t )\equiv \chitilde(\half + it )$  verses its $n=1$ approximation $\chitilde_1$  in \eqref{n1} as a function of $t$ in the region covering the first 2 zeros,
$t_1 = 14.13..$,  $t_2 = 21.02..$. (The orange curve is the approximation $\chitilde_1$.) }
 \label{EiApprox}
\end{figure}

\begin{figure}[b]
\centering\includegraphics[width=.35\textwidth]{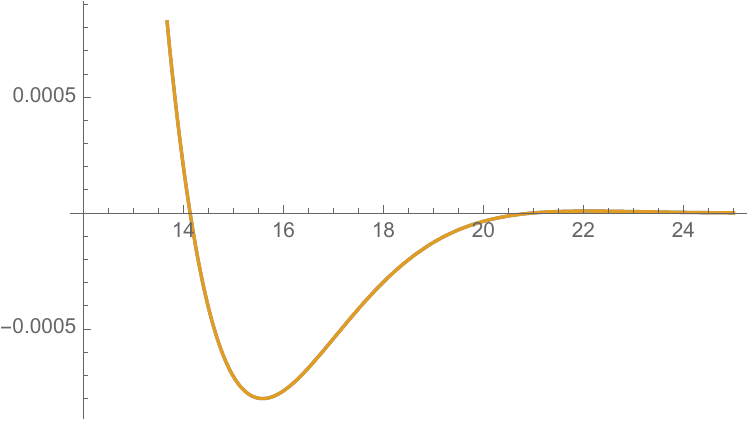}
\caption{A plot of  the exact $\chitilde (t )$  verses the approximation  $\chitilde (t) \approx \chitilde_1 (t) + \chitilde_2 (t)$  as a function of $t$ in the region covering the first 2 zeros,
$t_1 = 14.13..$,  $t_2 = 21.02$.  (The two plots are visually indistinguishable.) }
 \label{Upto2}
\end{figure}

Furthermore,  the integral in \eqref{FFTT}  can be performed for each individual  term in the sum over $n$ in \eqref{Phi} 
\beq
\label{chins}
\chitilde (t) = \sum_{n=1}^\infty  \chitilde_n (t), 
\eeq
where each $\chitilde_n$ can  expressed in terms of
the exponential integral function 
\beq
\label{ExpInt}
\Ei_\nu (r) = \int_1^\infty dx \, e^{-rx} x^{-\nu} .
\eeq
The result  is 
\beq
\label{n1}
\chitilde_n (t)  = 2 \, n^2 \pi \( 2 e^{-n^2 \pi} - \Re \[(\half + i t ) \, \Ei_{-\tfrac{1}{4}  +  \tfrac{it}{2} } (n^2 \pi)\] \).
\eeq
In Figure \ref{EiApprox} we plot the  approximation  $\chitilde_1 (t)$ against the exact $\chitilde (t)$ for low  $t$ up to the second zero 
$t_2 = 21..02$.  
One can see that this approximation gives reasonable results for the lowest zero $t_1$,   namely $t_1 \approx 14.06$ compared to the exact
$t_1 = 14.13...$. However it is clear that  one needs to include 
higher $n>1$ terms to better approximate the higher zeros,  as shown in Figure   \ref{Upto2} where we simply include $\chitilde_2$.    
With just  the simple approximation $\chitilde (t) \approx  \chitilde_1 (t) + \chitilde_2 (t)$,    the first zero is correct to  9 digits,  namely 
$t_1 \approx  14.13472510...$ verses the exact $t_1 =  14.13472514...$.   The second zero is correct to $7$ digits,  i.e. $t_2 = 21.022042$
verses the exact $t_2 = 21.022039$.   

Although one needs to include higher $\chitilde_{n>1}$ terms for higher zeros,   it  still turns out to be  interesting to consider the large $t$ asymptotics of  just the $n=1$ approximation in \eqref{n1}.      A somewhat crude approximation (good for $t  \gtrsim 10$) gives 
\beq
\label{EiAsymptotic}
 \, \chitilde_1 (t) \approx  -  \pi^{1/4} (2 t )^{3/2} \, e^{- \pi t/4} \, \cos \( \tfrac{t}{2} \log \(\tfrac{t}{2 \pi e} \) \),   ~~~~~~~~~~ (t \gtrsim 10).
\eeq
First note  that the $n=1$ approximation $\chitilde_1$ is a RAF in  the large $t$ limit due to the cosine factor in \eqref{EiAsymptotic}
and the fact that $t \log (t/2 \pi e)$ is monotonic for $t> 2 \pi e$.        
The Riemann zeros $t_n$ in this approximation arise from the cosine and satisfy
\beq
\label{AsymZeros}
\tfrac{t_n}{2} \log \( \tfrac{t_n}{2 \pi e} \) = (n - \tfrac{3}{2} ) \pi ~~~~~\Longrightarrow ~~ 
t_n = \frac{ (2n-3) \pi}{W( (2n-3)/2e )}, ~~~~~~(n\geq 1).
\eeq
Also note that for large $n$,   $t_n \approx 2 \pi n/\log n$,  in agreement with \eqref{tnLambert}. 
One problem with this large $t$ asymptotic is that if one sums over $n$ after taking the $t \to \infty$ limit,  the sum does not converge,  
indicating that integration over $u$ and the sum over $n$ do not commute in this large $t$ limit.

Let us close this section with a remark.
It  seems that it is only the existence of an infinite product 
 representation that makes $\chi (t) $ a {\em regular} alternating function, i.e. putting all minima or maxima at the right positions. 
More precisely,   the Euler product renders  $\lim_{\delta \to 0^+}  \theta(\half + \delta + i t) $ well defined and monotonic.
    This scenario seems to be well confirmed by the regular alternating nature observed in our numerical studies for the Dirichlet functions $L(s,\character)$  (see, for instance, Figure \ref{ChiRegAlternating} for the $\zeta (s)$ case).    In the next section we consider the  example of 
 the Davenport-Heilbronn $L$-function where there are zeros off the line and the regular alternating property is spoiled (see, later, Figure \ref{Davenportchi}).

\section{The counter example of Davenport-Heilbronn}

In this section we discuss the $L$-function of Davenport-Heilbronn (DH) \cite{DavenportH},   which satisfies a functional equation like the completed Riemann zeta but is known to have zeros off the critical  line. 
It is an interesting example in the present context since some of the above reasoning  for $\zeta (s)$ and $L(s, \character)$  applies,   thus it can be insightful 
  to understand how the RH fails in this case.     

Let $L(s, \character)$ denote the Dirichlet $L$-function based on this  mod $q=5$ character 
\beq
\label{chis}
\{\character(1), \character(2), \ldots, \character (5)\} = \{ 1, i, -i, -1, 0 \},   
\eeq
and defined as in \eqref{EPFchi}.  
The DH function $\CD (s)$ is engineered to satisfy a duality functional equation.  It is defined by the linear combination 
\beq
\label{DH}
\CD (s) \equiv  \tfrac{( 1- i \kappa)}{2} \, L(s,  \character ) +  \tfrac{( 1 + i \kappa)}{2} \, L(s,  \bar{\character}) 
\eeq
where $\kappa \equiv 
\tfrac{ \sqrt{10 - 2 \sqrt{5}} -2}{\sqrt{5} -1 }$.
For the remainder of this section we use the same notation $\chi (s)$,   $\theta(\sigma, t)$ and $\vartheta (\sigma,t)$ as above for $\zeta$,  however they all refer to the
DH function.  
Define it's completion 
\beq
\label{xiDH}
\xi (s) \equiv \( \tfrac{\pi}{5}  \)^{-s/2} \Gamma \( \tfrac{1+s}{2} \) \, \CD (s) .
\eeq
One can show that it satisfies the duality relation 
\beq
\label{DHfunc}
\xi (s) = \xi (1-s). 
\eeq

\def\rhostar{\rho_*}

As for $\zeta (s)$ we consider its  argument: 
\beq
\label{thetaDS} 
\theta (\sigma, t) = \arg \, \chi  (s) =  \thetaRS (\sigma, t)  +   \arg \, \CD (s)
\eeq
where 
\beq
\thetaRS  (\sigma,t)) = \arg\,  \Gamma \(  \tfrac{s+1}{2}  \) - \tfrac{t}{2} \log ( \tfrac{\pi}{5}) . 
\eeq
The appropriate Riemann-Siegel $\thetaRS$  on the critical line is
$\thetaRS (t) \equiv \thetaRS(\half, t)$.  
The exact counting function obtained from the Cauchy argument principle,  which also knows about any potential zeros off the line since it counts all zeros in the strip,  is
\beq
\label{NoftDH}
N(T) =\lim_{\delta \to 0^+}  \theta (\half + \delta, T) /\pi . 
\eeq

The analog of eqn.  \eqref{Horzn}  for the hypothetical $n$-th zero of the DH function on the critical line is simply
\beq
\label{transDH}
\lim_{\delta \to 0^+} \theta(\half+\delta, t_n) =  \( n- \tfrac{1}{2} \) \pi .
\eeq
Indeed one can easily check numerically that \eqref{transDH} correctly gives the first $43$ zeros,  which are all on the critical line.   
However for $n = 44$ and $45$ there is no solution to the equation \eqref{transDH} and this signifies two zeros off the line which are complex conjugates.    They are: 
\beq
\label{DHoffLine} 
\CD (\rhostar) = 0 ~~~{\rm for} ~~ \rhostar = 0.8085171825 ~\pm ~ i \, 85.6993484854.
\eeq
There are an infinite number of zeros on the line which satisfy \eqref{transDH},   and also an infinite number of zeros off the line where there are no solutions to \eqref{transDH}.  
The next $n$ where the latter occurs is $n=63$.     

\begin{figure}[b]
\centering\includegraphics[width=.5\textwidth]{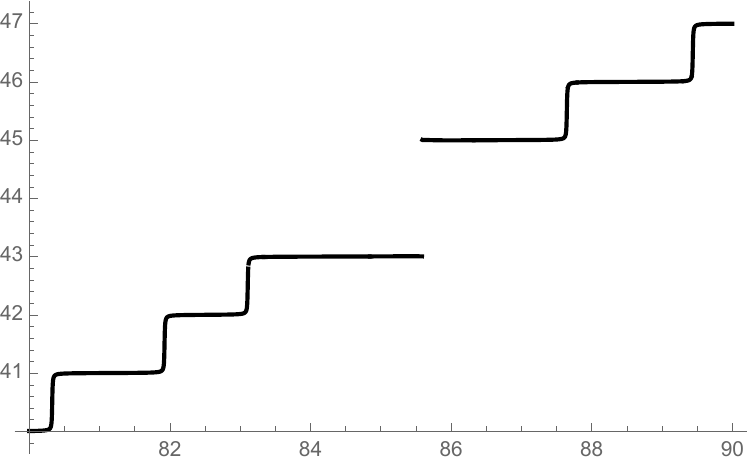}
\caption{A plot of $N(t)$ in \eqref{NoftDH} as a function of $t$ in the range of the first zero off the line.  We emphasize that 
the non-zero $\delta$ in \eqref{NoftDH} is necessary otherwise $N(t)$ would jump {\it discontinuously} by $1$ at zeros on the line.}
 \label{DHNofT}
\end{figure}

Let us attempt to interpret the above results in the context of our scattering description of the zeros on the line in order to understand how the RH fails.   
There is one reasonably simple answer.    Each $L(s, \character)$ function in the definition \eqref{DH} satisfies an Euler product formula in \eqref{EPFchi},   
however the linear combination in $\CD (s)$ does not have such a representation.   Thus the scattering problem  for a {\em finite} number $N$ of impurities, as defined in Sections III and IV for $\zeta (s)$, cannot even be defined for the Davenport-Heilbronn $L$-function, since for the latter  there only exists the Bethe Ansatz equation for the infinite thermodynamic limit directly,  specifically by analytic continuation.   In the absence of any finite $N$ formulation of the Bethe Ansatz equations, which guarantees at least for any finite
 $N$ the continuity of the sum of the phase-shifts coming from the scattering of the impurities, it may happen that the argument of $\CD (s)$ on the critical line is simply not well-defined. For the DH function this happens where potential zeros on the line disappear and are manifested as zeros off the line.       
One can clearly see that the left hand side of \eqref{transDH} is discontinuous near the zero off the line where it jumps by $2 \pi$,  which is why there is no solution 
(see Figure \ref{DHNofT}). 
We emphasize that the non-zero $\delta$ in \eqref{NoftDH} is necessary otherwise $N(t)$ would also jump {\it discontinuously} by $1$ at zeros on the line.
We argued that this does not occur for any finite value of $N$ of the Bethe Ansatz equations associated to the Riemann $\zeta$-function. For the Dirichlet $L$-functions of primitive non-principal characters,  we argued  that the argument of these functions is continuous along the critical line based on  the well-behaved convergence properties of the relevant  truncated series.  
d{figure}

\begin{figure}[t]
\centering\includegraphics[width=.4\textwidth]{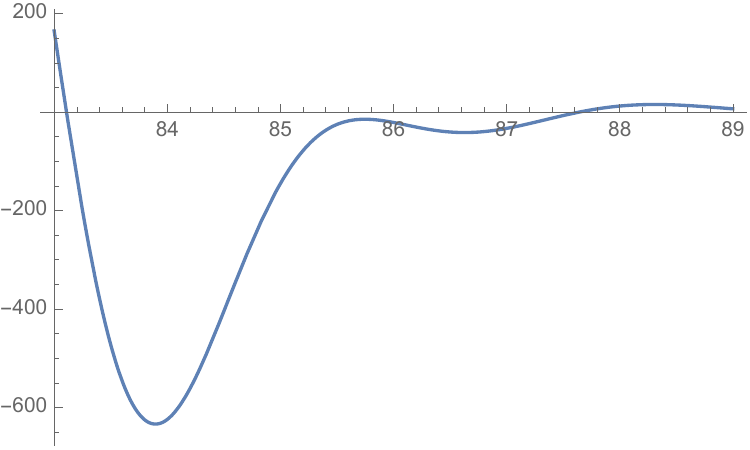}
\caption{A plot of $\chi (\half + i t )$ for the Davenport-Heilbronn $L$-function  a function of $t$ in the region around the first zero off the line in \eqref{rhozero}.   (Magnified by $10^{30}$.)}
 \label{Davenportchi}
\end{figure}

Finally there is also a signature of the zeros off the line in the function $\chi (\half + i t)$ for $t$ near such zeros.   
One can  check numerically that the regular alternating property displayed in Figure \ref{ChiRegAlternating} is violated where there are zeros off the critical line, see Figure \ref{Davenportchi}, with a pattern which  essentially displays the same phenomenon as  the one summarized by Figure \ref{zerooooo}, i.e. one maximum has gone below the $t$ axis, so that two real zeros have disappeared, becoming instead complex.

\section{Conclusions}

The scattering problem we constructed which leads exactly to the Riemann zeros on the critical line requires {\it both} the duality equation 
$\chi (s) = \chi (1-s)$ and the Euler product formula. The obvious question is: how could be that the Bethe Ansatz equations fail in finding {\em all} real solutions 
such that the GRH is false?   We have argued that it is more and more likely that there is a unique solution 
to the exact equation \eqref{FLeqn} in the limit of large $n$ since the fluctuating $\CS (t)$ term is more and more subleading as $t\to \infty$.     
The only way we can imagine that the completeness of the Bethe ansatz system fails is if $\CS  (t)$ becomes somehow ill-defined in some region of $t$. Specifically  if $\CS (t)$ discontinuously jumps by $2$,  there would be no solution to \eqref{FLeqn} for some $n$, and this would signify a pair of zeros off the line  which are complex conjugates of each other,  or a double zero on the line. 
Indeed this is what occurs for the Davenport-Heilbronn counterexample discussed in Section VII, i.e. a function which satisfies the duality relation but which does not have an infinite product representation.
 But, for a phase shift coming from the Riemann zeta  function, we have its Euler product representation. This ensures its continuity for any arbitrary truncation in its number of terms. Moreover, Selberg's central limit theorem for $\CS (t)$ is based on the truncation of the Euler product.  
As a matter of fact, we have shown that adding more terms to the Euler product representation only increases the accuracy of computing the actual zeros $t_n$, never causing them to disappear,  so long as one truncates the product properly,  namely $N <  t^2$ as discussed above. The situation is even better for the Dirichlet $L$-functions based on non-principal characters,  where there is convergence for the  infinite product to the right of the critical line \cite{LMDirichlet}. For these functions there are even stronger  reasons to believe that the Bethe ansatz equations are complete and therefore that all their non-trivial zeros are along the critical axis $\Re (s) = \half$. 
We have also argued that the RH is true if $\chi(\half + i t)$ is a regular alternating function  (RAF) of $t$ (see above for the definition  of  RAF), and have provided some evidence for this in Section VI. In conclusion, viewing the non-trivial zeros of the Riemann or Dirichlet $L$-functions as roots of a Bethe Ansatz system of equations for an integrable scattering model gives a different perspective on the validity of the (Generalised) Riemann Hypothesis and, at the same time, clarifies the important role played by the infinite-product representation of these functions. It will be interesting to investigate further the aspects emerged from the analysis of this paper, in particular the scattering problem associated to phase shifts coming from a random sequence of integers.     

\bigskip
\bigskip

\section{Acknowledgements} 

We thank  German Sierra and Ghaith Hiary for discussions.   
AL would like to thank SISSA where this work was started in  June 2023.

\end{document}